\documentclass[journal]{IEEEtran}
\usepackage{amsfonts}
\IEEEoverridecommandlockouts

\ifCLASSINFOpdf
\else
\fi
\usepackage{epstopdf}
\usepackage{multirow}
\usepackage{cite}
\usepackage{stfloats}
\usepackage{color}
\usepackage{epsfig}
\usepackage{graphicx}
\usepackage{psfig}
\usepackage{epsf}
\usepackage{slashbox}
\usepackage{float}
\usepackage{amssymb }
\usepackage{cite}
\usepackage{stfloats}
\usepackage{color}
\usepackage{epsfig}
\usepackage{graphicx}
\usepackage{psfig}
\usepackage{epsf}
\usepackage{slashbox}
\usepackage{float}
\usepackage[cmex10]{amsmath}
\usepackage{algorithm}
\usepackage{algorithmic}
\usepackage{subfigure}

\begin{document}
\title{Performance Analysis in DF Based Cooperative SWIPT Networks with Direct Link} %Based SWIPT with Dual-Hop DF Relaying in the Presence of a Direct Link
%Hybrid Scheme for SWIPT Based DF Relay Networks with Direct Link
\author{\IEEEauthorblockN{Yingting Liu, Yinghui Ye, Haiyang Ding, Jianmei Shen, and Hongwu Yang}
\thanks{This work was supported in part by the National Natural Science Foundation of China (Grant No. 61861041, 61871387 and 11664036), and in part by the National Key Research and Development Plan of China (Grant No. 2018YFE0100500).}% <-this % stops a space
\thanks{Yingting Liu, Jianmei Shen and Hongwu Yang are with the College of Physics and Electronic Engineering, Northwest Normal University, Lanzhou, 730070, China.
}
\thanks{Yinghui Ye and Haiyang Ding are with the Integrated Service Networks Lab of Xidian University, Xi'an, China. Haiyang Ding is also with School of Information and Communications, National University of Defense Technology, Xi'an, China. The corresponding author is Yingting Liu (e-mail: liuyt2018@163.com)}
}% the National Key Research and Development Program of China (2016YFB1200202), the Natural Science Foundation of Shaanxi Province (2017JZ022), the Natural Science Foundation of China (61501371), the 111 Project of China (B08038).
%\markboth{IEEE WIRELESS COMMUNICATIONS LETTERS, No. XX, MONTH YY, YEAR 2019}
%{\MakeLowercase{\textit{et al.}}: }
\maketitle

\begin{abstract}
%In the paper, a simultaneous wireless information and power information transfer (SWIPT) system consists of one source node (S), one relay node (R) and one destination node (D) is proposed. The system adopts the power splitting protocol (PSR). Each transmission is divided into two equal time slots. In the first time slot, S transmits the signal to the R and D. R harvests energy using a dynamic power splitter adjusting with the instantaneous channel state information (CSI) in the first time slot, and decodes and forwards the information to D in the second time slot, respectively. We first derive the optimal dynamic power splittingF factor resulting in maximal achievable rate, and then derive the optimal ergodic capacity of the system based on it. The accurate analytical integral expression and approximate analytical closed form expression for the ergodic capacity is obtained. Numerical results verify the analytical results and show that the proposed scheme has the better performance compared to the existing fixed power splitter scheme with a direct link and dynamic power splitter scheme without a direct link.
This letter proposes a dynamic power splitting scheme (DPSS) for decode-and-forward (DF) based cooperative simultaneous wireless information and power transfer (SWIPT) networks with direct link. The relay node adopts an optimal dynamic power splitting factor determined by instantaneous channel state information (CSI) to harvest energy and process information. The expressions for the optimal dynamic power splitting factor, outage probability and ergodic capacity of the proposed network are derived. Numerical results show that the proposed scheme is better than or the same as the existing PS schemes in terms of outage probability, while it achieves higher ergodic capacity compared to the existing PS schemes.

\end{abstract}

%\begin{IEEEkeywords}
%Simultaneous wireless information and power information transfer, dynamic power splitting factor, decode-and-forward, outage probability, ergodic capacity, direct link.
%\end{IEEEkeywords}
\IEEEpeerreviewmaketitle
\section{Introduction}
%\IEEEPARstart{I}{n} conventional communication systems, all devices use constant power, such as batteries or electricity, to drive their operation. All communication nodes consume huge electric energy every year. In recent years, energy harvesting (EH) technology appeals to the interest of the researchers, because the devices adopt the EH technology can harvest energy from the surrounding environment. This will prolong the lifetime of communication devices and decrease the carbon emission of the earth [1]. Different from the discontinuity performance of conventional EH technologies, such as solar, wind and thermoelectric effects, simultaneous wireless power and information transfer (SWIPT) can realize the information transmission as well as power transfer simultaneously utilizing the ability of radio frequency (RF) carrying energy. Due to the fact that wireless devices are ubiquitous nowadays, SWIPT can supply inexhaustible energy for wireless devices and obtains more and more attention in the literature [2].
\IEEEPARstart{C}{ompared} to the non-cooperative network, cooperative communication can obviously improve system performance. Radio frequency (RF) signals have the ability of carrying energy as well as carrying information at the same time. Using the ability of RF signals, simultaneous wireless information and power transfer (SWIPT) based cooperative communication network has attracted much attention in the literatures, such as [1$-$6] and references therein.

The authors studied outage probability, ergodic capacity and symbol error rate in one-way [2$-$3] or two-way [4$-$6] network. In above literatures, the relay node can use power splitting (PS) scheme and time switching (TS) scheme to harvest energy, and adopt amplify-and-forward (AF) or decode-and-forward (DF) mode to forward signals. In a PS scheme, the received power is divided into two portions in proportion for harvesting energy and processing information simultaneously. This results in a better performance for the PS scheme by comparing to the TS scheme, and thus we focus on the design of PS in this letter

Until now, the design of PS scheme is based on the statistic or instantaneous channel state information (CSI).  We refer the statistic CSI  based PS scheme as a fixed power splitting scheme (FPSS), where the PS factor is constant over all transmission blocks if the statistic CSI remains unchanged [2$-$5]. On the contrary, we terms the instantaneous CSI based PS as the dynamic power splitting scheme (DPSS), in which the PS factor is constant over a transmission block and may be changed in the next block. The authors in [6, 10] proposed a DPSS, in which the power splitter adjusts its parameter in each transmission based on the instantaneous CSI and the target data rate. As shown in [7], in the cooperative communication, the direct link (DL) between the source node and the destination node can usually supply the additional diversity gain that can enhance the performance of the network. In [2$-$6], the authors neglected the DL from the source to the destination. The authors in [8] analyzed the average symbol error rate for TS based one-way DF networks with or without DL in Nakagami-$m$ fading channels. The authors in [9] and [10] studied the outage probability of one-way PS based SWIPT network with DL. In [9], the relay adopts FPSS and AF protocol, and in [10], the relay adopts DPSS and DF protocol, respectively. The authors in [11] presented a novel DPSS, in which the relay adjusts the PS factor aiming to maximizing the end-to-end signal to noise ratio (SNR). As described in [11], the existing DPSS for DF based SWIPT network without DL [6, 10] is able to minimize the outage probability but  fails to maximize the ergodic capacity. However, the authors do not consider the effect of the DL. Accordingly, an proper DPSS scheme is required for DF based cooperative SWIPT networks with DL and this motivates this work.

Our contributions of this letter are summarized as follows.
\begin{itemize}
\item We propose a novel DPSS for DF based SWIPT network with DL. We derive the closed-form expression of the optimal dynamic PS factor.
\item The approximate analytical expressions of the outage probability and ergodic capacity are derived. The simulations verify our derived results and show that the proposed scheme can minimize the outage probability and maximize the ergodic capacity at the same time compared to the existing schemes.
\end{itemize}
\vspace{-15pt}
\section{System Model}
%\begin{figure}[htbp]
%\centering
%\includegraphics[width=3 in]{systemmodel}
%\caption{The system model.}
%\label{Fig.1}
%\end{figure}
In the proposed DF based cooperative SWIPT network, the source node (denoted by $S$) transmits its information to the destination node (denoted by $D$) via the help of an energy constrained relay node (denoted by $R$). Each transmission $T$ can be divided into two equal time duration slots $\frac{T}{2}$. The channel coefficients of the corresponding links $S\rightarrow D$, $S\rightarrow R $ and $R\rightarrow D$ are denoted by $h_0$, $h_1$ and $h_2$, respectively. The channels are modeled as quasi-static, which means the channels keep constant over each transmission but may vary in different transmissions. The channel fading coefficients are assumed to follow Rayleigh distributions and the means of the channel gains $|h_{0}|^{2}$, $|h_{1}|^{2}$ and $|h_{2}|^{2}$ are denoted by $\frac{1}{\lambda_0}$, $\frac{1}{\lambda_1}$ and $\frac{1}{\lambda_2}$. It is assumed that all node are equipped with single antenna [2$-$6] and the instantaneous CSI can be obtained by channel estimation [6, 10$-$11].

In the first time slot $\frac{T}{2}$, $S$ sends the signal to $R$ and $D$, and the received signals for processing information at $R$ and $D$ can be written as
\begin{equation}\label{1}
y_{r}(t)=\sqrt{(1-\rho)P_{s}}h_{1}s(t)+n_{r}(t),
\end{equation}
\begin{equation}\label{2}
y_{sd}(t)=\sqrt{P_{s}}h_{0}s(t)+n_{d}(t),
\end{equation}
where $s(t)$ denotes the transmitted signal by $S$ and has unit energy. $P_{s}$ denotes the transmitted power and $0 \le \rho  \le 1$ is the dynamic PS factor determined by instantaneous CSI, which will be detailed in the following. $n_{r}(t)$ and $n_{d}(t)$ denote the received additive white Gaussian noises (AWGNs) with zero means and the variances $\sigma_{r}^{2}$ and $\sigma_{d}^{2}$ at $R$ and $D$, respectively. $n_{r}(t)$ and $n_{d}(t)$ both consist of the AWGN introduced by the receiving antenna and the AWGN produced by the RF band to baseband conversion. For simplicity, it is assumed that $\sigma^{2}=\sigma_{r}^{2}=\sigma_{d}^{2}$, as in [6, 9$-$11].

Based on (1) and (2), the SNR in the first slot at $R$ and $D$ can be derived as $\gamma_{r}=(1-\rho)\gamma_{\rm{in}} |h_{1}|^{2}$ and $\gamma_{sd}=\gamma_{\rm{in}} |h_{0}|^{2}$, respectively, where $\gamma_{\rm{in}}=\frac{P_{s}}{\sigma^{2}}$ denotes the SNR transmitted by $S$.
The harvested energy at $R$ is written as $E_{r}=\eta\rho P_{s}|h_{1}|^{2}\frac{T}{2}$. The transmitted power by $R$ in the second time slot can be derived as $P_{r}=\eta\rho P_{s}|h_{1}|^{2}$, where $0<\eta<1$ is the energy conversion efficiency.
The received signal in the second slot at $D$ can be derived as
\begin{equation}\label{3}
y_{rd}(t)=\sqrt{\eta\rho P_{s}|h_{1}|^{2}}h_{2}s(t)+n_{d}(t).
\end{equation}
And the corresponding received SNR can be expressed as $\gamma_{rd}=\eta\rho\gamma_{\rm{in}} |h_{1}|^{2}|h_{2}|^{2}$.

\section{PS Scheme Design and Performance Analysis}
Adopting the maximal ratio combining (MRC), the achievable data rate at $D$ is given as [7, eq. (15)]
\begin{equation}\label{4}
R=\frac{1}{2}\min\{\log_{2}(1+\gamma_{r}),\log_{2}(1+\gamma _d)\},
\end{equation}
where ${\gamma _d} = \gamma _{sd} + \gamma _{rd}$.
The problem to solve the maximal achievable rate at $D$ is equivalent to following problem, given by
\begin{equation}\label{5}
{\gamma_{\rm{op}}} = \mathop {\max \min }\limits_{_{0 \le \rho  \le 1}} \left( {{\gamma _r},{\gamma _d}} \right).
\end{equation}

Observing the expressions of $\gamma_{r}$, $\gamma_{sd}$ and $\gamma_{rd}$, it is found that $\gamma_{r}$ and  $\gamma_{d}$ are the decreasing and increasing functions with respect to the dynamic PS factor $\rho$, respectively.

%\begin{figure}[htbp]
%\centering
%\includegraphics[width=2 in]{fig1inthepaper}
%\caption{Illustration of optimal power splitting factor.}
%\label{Fig.2}
%\end{figure}
%\begin{lemma} \label{lemma1}
\emph{Lemma} 1: The optimal $\rho^{\ast}$ leading to maximal achievable rate can be derived as
\begin{equation}\label{6}
{\rho ^ * } = \left\{ \begin{array}{l}
~~~~~~~~~~~0,~~~~~~~~~~~|{h_1}{|^2} < |{h_0}{|^2}\\
\frac{|h_{1}|^{2}-|h_{0}|^{2}}{|h_{1}|^{2}+\eta|h_{1}|^{2}|h_{2}|^{2}},~~~~~|{h_1}{|^2} \geq|{h_0}{|^2}
\end{array} \right.
\end{equation}
%\end{lemma}

\emph{Proof}: When $|h_{1}|^{2}<|h_{0}|^{2}$, the curve of $\gamma_{r}$ has no crossing point with respect to that of $\gamma_{d}$, and the optimal received SNR at $D$ is achievable when $\rho^{\ast}=0$ and equal to $\gamma_{\rm{op}}^1=\gamma_{\rm{in}}|h_{0}|^{2}$. When $|h_{1}|^{2}\geq|h_{0}|^{2}$, the optimal received SNR at $D$ is achievable by equating $\gamma_{r}$ to $\gamma_{d}$. In this case, $0\leq\rho^{\ast}=\frac{|h_{1}|^{2}-|h_{0}|^{2}}{|h_{1}|^{2}+\eta|h_{1}|^{2}|h_{2}|^{2}}<1$. Substituting the value of $\rho^{\ast}$ into (5), the corresponding maximal SNR can be obtained as $\gamma_{\rm{op}}^2=\gamma_{\rm{in}}\frac{|h_{0}|^{2}+\eta|h_{1}|^{2}|h_{2}|^{2}}{1+\eta|h_{2}|^{2}}$.

\emph{Remark}: As exhibited in [7, eq. (15)] or (4),  the achievable rate is determined by the minimum of $\gamma_{r}$ and $\gamma _{sd} + \gamma _{rd}$. This  is valid in conventional cooperative communication networks, since all nodes are powered by a battery and all the SNRs, i.e., $\gamma_{r}$, $\gamma_{sd}$ and $\gamma_{rd}$, are larger than zero.
However, eq. (4) may be invaild in our considered  DF based cooperative SWIPT network. This is because the PS based relay is powered by the RF signals from the source and the transmit power of such relay may equal zero. For example, when $|h_{1}|^{2}<|h_{0}|^{2}$, we have $\rho ^ * =0$ and $\gamma_{rd}=\eta\rho ^ *\gamma_{\rm{in}} |h_{1}|^{2}|h_{2}|^{2}=0$. In such case,  the achievable rate by (4) is  $\frac{{\rm{1}}}{{\rm{2}}}{\log _2}\left( {1 + {\gamma _{{\rm{in}}}}{{\left| {{h_{\rm{1}}}} \right|}^{\rm{2}}}} \right)$, which is smaller than that achieved by the DL link $\frac{{\rm{1}}}{{\rm{2}}}{\log _2}\left( {1 + {\gamma _{{\rm{in}}}}{{\left| {{h_0}} \right|}^{\rm{2}}}} \right)$.
In this case, the proposed model is equivalent to the non-cooperative  transmission and we can obtain $\gamma_{\rm{op}}^1=\gamma_{\rm{in}}|h_{0}|^{2}$.
%Submitting $\rho^*=0$ into the modified expression, we have
%In this case, the proposed model is equivalent to the non-cooperative  transmission and we can obtain $\gamma_{\rm{op}}^1=\gamma_{\rm{in}}|h_{0}|^{2}$.
\vspace{-15pt}
\subsection{Outage Probability Analysis}
The outage probability is defined as the probability that the achievable data rate is below the target data rate $R_{\rm{th}}$. The outage probability of the proposed network can be written as
\begin{align}\label{7}
P_{\rm{out}}=P\left( {{\gamma _{{\rm{op}}}} < {\gamma _{{\rm{th}}}}} \right) = {P_1} + {P_2},
\end{align}
where $\gamma_{\rm{th}}=2^{2R_{\rm{th}}-1}$, ${P_1} = P\left( {\gamma _{{\rm{op}}}^1 < {\gamma _{{\rm{th}}}},{\rho ^*} = 0} \right)$ and ${P_2} = P\left( {\gamma _{{\rm{op}}}^2 < {\gamma _{{\rm{th}}}},{\rho ^*} = \frac{{|{h_1}{|^2} - |{h_0}{|^2}}}{{|{h_1}{|^2} + \eta |{h_1}{|^2}|{h_2}{|^2}}}} \right)$.

Let $X=|h_{0}|^2$, $Y=|h_{1}|^2$ and $Z=|h_{2}|^2$, so the corresponding probability destiny functions (PDFs) of the variables $X$, $Y$ and $Z$ can be written as ${f_X}\left( x \right) = {\lambda _0}{e^{ - {\lambda _0}x}}$, ${f_Y}\left( y \right) = {\lambda _1}{e^{ - {\lambda _1}y}}$ and ${f_Z}\left( z \right) = {\lambda _2}{e^{ - {\lambda _2}z}}$.

Accordingly, $P_{1}$ can be calculated as
\begin{align}\label{8}
{P_1} &= P\left( {{\gamma _{{\rm{in}}}}|{h_0}{|^2} < {\gamma _{{\rm{th}}}},|{h_1}{|^2} < |{h_0}{|^2}} \right)\notag\\
&= P\left( {x < a,x > y} \right) = \int_0^a {\int_y^a {{\lambda _0}{e^{ - {\lambda _0}x}}{\lambda _1}{e^{ - {\lambda _1}y}}dxdy} } \notag\\
&= \int_0^a {\left( {{e^{ - {\lambda _0}y}} - {e^{ - {\lambda _0}a}}} \right)} {\lambda _1}{e^{ - {\lambda _1}y}}dy\notag\\
&= \frac{{{\lambda _1}}}{{{\lambda _0} + {\lambda _1}}}\left( {1 - {e^{ - \left( {{\lambda _0} + {\lambda _1}} \right)a}}} \right) - {e^{ - {\lambda _0}a}}\left( {1 - {e^{ - {\lambda _1}a}}} \right),
\end{align}
where $a=\frac{\gamma_{\rm{th}}}{\gamma_{\rm{in}}}$.

On the other hand, $P_{2}$ can be calculated as
\begin{align}\label{9}
{P_2} &= P\left( {{\gamma _{{\rm{in}}}}\frac{{|{h_0}{|^2} + \eta |{h_1}{|^2}|{h_2}{|^2}}}{{1 + \eta |{h_2}{|^2}}} < {\gamma _{{\rm{th}}}},|{h_0}{|^2} \leq |{h_1}{|^2}} \right)\notag\\
& = P\left( {\left( {\eta y - \eta a} \right)z < a - x,x \leq y} \right).
\end{align}
Considering the fact $X>0$, $Y>0$, $Z>0$ and the condition $X\leq Y$, and omitting the intermediate manipulation operations, $P_{2}$ can be further derived as
\begin{align}\label{10}
{P_2} &= \underbrace {\int_0^a {\int_0^y {{\lambda _0}{e^{ - {\lambda _0}x}}{\lambda _1}{e^{ - {\lambda _1}y}}dxdy} } }_{{P_{21}}\left( {0 < x < y < a} \right)} \notag\\
&+ \underbrace {\int_a^\infty {\int_0^a {\left( {1 - {e^{ - {\lambda _2}\left( {\frac{{a - x}}{{\eta y - \eta a}}} \right)}}} \right){\lambda _0}{e^{ - {\lambda _0}x}}{\lambda _1}{e^{ - {\lambda _1}y}}dxdy} } }_{{P_{22}}\left( {x < a,y > a} \right)}.
\end{align}
In (10), the first term $P_{21}$ can be calculated as
\begin{align}\label{11}
{P_{21}} = \frac{{{\lambda _0}}}{{{\lambda _0} + {\lambda _1}}} + \frac{{{\lambda _1}}}{{{\lambda _0} + {\lambda _1}}}{e^{ - \left( {{\lambda _0} + {\lambda _1}} \right)a}} - {e^{ - {\lambda _1}a}}.
\end{align}
The second term $P_{22}$ can be calculated as
\begin{small}
\begin{align}\label{12}
{P_{22}} = {P_{221}} - {\lambda _0}{\lambda _1}\underbrace {\int_a^\infty  {\int_0^a {\left( {{e^{ - {\lambda _2}\left( {\frac{{a - x}}{{\eta y - \eta a}}} \right)}}} \right){e^{ - {\lambda _0}x}}{e^{ - {\lambda _1}y}}dxdy} } }_{{P_{222}}},
\end{align}
\end{small}where ${P_{221}}=\left( {1 - {e^{ - {\lambda _0}a}}} \right){e^{ - {\lambda _1}a}}$.

Invoking the change of variables $u=y-a$, $P_{222}$ can be derived as
\begin{small}
\begin{align}\label{13}
&{P_{222}} = {e^{ - {\lambda _1}a}}\int_0^a {{e^{ - {\lambda _0}x}}} \int_0^\infty  {{e^{ - {\lambda _1}u - {\lambda _2}\frac{{a - x}}{{\eta u}}}}dudx}\notag\\
& = \frac{{{e^{ - {\lambda _1}a}}}}{{{\lambda _1}}}\int_0^a {{e^{ - {\lambda _0}x}}\sqrt {\frac{{4\left( {a - x} \right){\lambda _1}{\lambda _2}}}{\eta }} } {K_1}\left( {\sqrt {\frac{{4\left( {a - x} \right){\lambda _1}{\lambda _2}}}{\eta }} } \right)dx\\
&\approx \frac{{{e^{ - {\lambda _1}a}}}}{{{\lambda _1}}}\int_0^a {{e^{ - {\lambda _0}x}}} \left( {1 + \frac{{\left( {a - x} \right){\lambda _1}{\lambda _2}}}{\eta }\ln \left( {\frac{{\left( {a - x} \right){\lambda _1}{\lambda _2}}}{\eta }} \right)} \right)dx\\
 & = \frac{{{e^{ - {\lambda _1}a}}}}{{{\lambda _0}{\lambda _1}}}\left( {1 - {e^{ - {\lambda _0}a}}} \right) + \Phi \left( {a{e^{{\lambda _0}a}} - \frac{1}{{{\lambda _0}}}\left( {{e^{{\lambda _0}a}} - 1} \right)} \right)\notag\\
&+ \Lambda \left( {a\ln a{e^{{\lambda _0}a}} - \frac{1}{{{\lambda _0}}}\left( {{e^{{\lambda _0}a}} - 1} \right) - \int_0^a {{e^{{\lambda _0}x}}\ln xdx} } \right).
\end{align}
\end{small}In (13), we use the integration $\int_0^\infty  {{e^{ - \frac{\beta }{{4x}} - \gamma x}}dx = } \sqrt {\frac{\beta }{\gamma }} {K_1}\left( {\sqrt {\beta \gamma } } \right)$ [12, 3.324.1] and in (14), we utilize the fact that $a\rightarrow 0$ in the high SNR regime and the equivalent infinitesimal replacement $\theta {K_1}\left( \theta  \right) \approx 1 + \frac{{{\theta ^2}}}{2}\ln \left( {\frac{\theta }{2}} \right)$ when $\theta\rightarrow0$ [6, 10]. ${K_1}\left( \cdot \right)$ is the the first order modified Bessel function of the second kind. In (15), $\Phi  = \frac{{{\lambda _2}{e^{ - \left( {{\lambda _0} + {\lambda _1}} \right)a}}\ln \left( {\frac{{{\lambda _1}{\lambda _2}}}{\eta }} \right)}}{{\eta {\lambda _0}}}$ and $\Lambda  = \frac{{{\lambda _2}{e^{ - \left( {{\lambda _0} + {\lambda _1}} \right)a}}}}{{\eta {\lambda _0}}}$.

To the best of our knowledge, the integration ${\int_0^a {{e^{{\lambda _0}x}}\ln xdx} }$  shown in the last term of (15) has no closed-form result. For obtaining the closed-form result, we can use the Gaussian-Chebyshev quadrature [10] to approximate the the integration as
\begin{equation}\label{16}
\int_0^a {{e^{{\lambda _0}x}}\ln xdx}  \approx \frac{a}{2}\omega \sum\limits_{i = 1}^N {\sqrt {1 - f_i^2} } {e^{\frac{{{\lambda _0}a}}{2}{c_i}}}\ln \left( {\frac{a}{2}{c_i}} \right),
\end{equation}
where ${\omega}=\frac{\pi}{N}$, $f_{i}=\cos(\frac{(2i-1)\pi}{2N})$ and ${c_{{i}}} = {f_{{i}}} + 1$. $N$ is the parameter determining the tradeoff between the complexity and the accuracy.

Substituting (8)$-$(16) into (7), the analytical expression of the proposed model for the outage probability can be obtained. For the limit of the page, it is neglected in the letter.

The diversity order of the proposed model can derived as $D =  - \mathop {\lim }\limits_{{\gamma _{{\rm{in}}}} \to \infty } \frac{{\log \left( {\frac{{\log \left( {{\gamma _{{\rm{in}}}}} \right)}}{{{\gamma _{{\rm{in}}}}^2}}} \right)}}{{\log \left( {{\gamma _{{\rm{in}}}}} \right)}} = 2 $.
%\vspace{-15pt}
%\begin{equation}
%D =  - \mathop {\lim }\limits_{{\gamma _{{\rm{in}}}} \to \infty } \frac{{\log \left( {\frac{{\log \left( {{\gamma _{{\rm{in}}}}} \right)}}{{{\gamma _{{\rm{in}}}}^2}}} \right)}}{{\log \left( {{\gamma _{{\rm{in}}}}} \right)}} = \mathop {\lim }\limits_{{\gamma _{{\rm{in}}}} \to \infty } \frac{{\log \left( {{\gamma _{{\rm{in}}}}^2} \right) - \log \log \left( {{\gamma _{{\rm{in}}}}} \right)}}{{\log \left( {{\gamma _{{\rm{in}}}}} \right)}} = 2 \notag.
%\end{equation}
%\begin{equation}
%D =  - \mathop {\lim }\limits_{{\gamma _{{\rm{in}}}} \to \infty } \frac{{\log \left( {\frac{{\log \left( {{\gamma _{{\rm{in}}}}} \right)}}{{{\gamma _{{\rm{in}}}}^2}}} \right)}}{{\log \left( {{\gamma _{{\rm{in}}}}} \right)}} = 2 \notag.
%\end{equation}
\vspace{-15pt}
\subsection{Ergodic Capacity Analysis}
The optimal ergodic capacity of the proposed model can be written as
\begin{align}\label{17}
C = {C_1} + {C_2},
\end{align}
where $C_{1}$ and $C_{2}$ are the ergodic capacities corresponding to the cases $\rho ^ * =0$ and $\rho ^ *=\frac{{|{h_1}{|^2} - |{h_0}{|^2}}}{{\eta |{h_1}{|^2}|{h_2}{|^2} + |{h_1}{|^2}}}$, respectively.

$C_{1}$ can be derived as (18)$-$(21), which are exhibited at the top of next page.
\newcounter{mytempeqncnt}
\begin{figure*}[t]
\normalsize
\setcounter{mytempeqncnt}{\value{equation}}
\setcounter{equation}{17}
\begin{small}
\begin{align}
{C_1} &= \frac{1}{2}\int_0^\infty  {\int_y^\infty  {{\lambda _0}{e^{ - {\lambda _0}x}}{\lambda _1}{e^{ - {\lambda _1}y}}{{\log }_2}\left( {1 + {\gamma _{{\rm{in}}}}x} \right)dxdy} } \\
&  = \frac{1}{{2\ln 2}}\int_0^\infty  {{\lambda _1}{e^{ - {\lambda _1}y}}\left( { - \ln \left( {1 + {\gamma _{{\rm{in}}}}x} \right){e^{ - {\lambda _0}x}}|_y^\infty  + \int_y^\infty  {\frac{{{\gamma _{{\rm{in}}}}{e^{ - {\lambda _0}x}}}}{{1 + {\gamma _{{\rm{in}}}}x}}} dx} \right)dy} \\
& =  - \frac{{{\lambda _1}}}{{2\left( {{\lambda _{\rm{0}}} + {\lambda _1}} \right)\ln 2}}{e^{\frac{{{\lambda _0} + {\lambda _1}}}{{{\gamma _{{\rm{in}}}}}}}}{{\rm{E}}_{\rm{i}}}\left( { - \frac{{{\lambda _0} + {\lambda _1}}}{{{\gamma _{{\rm{in}}}}}}} \right) - \frac{{{\lambda _1}{e^{\frac{{{\lambda _0}}}{{{\gamma _{{\rm{in}}}}}}}}}}{{2\ln 2}}\int_0^\infty  {{e^{ - {\lambda _1}y}}\left( {{{\rm{E}}_{\rm{i}}}\left( { - {\lambda _0}y - \frac{{{\lambda _0}}}{{{\gamma _{{\rm{in}}}}}}} \right)} \right)} dy\\
&\approx  - \frac{{{\lambda _1}}}{{2\left( {{\lambda _{\rm{0}}} + {\lambda _1}} \right)\ln 2}}{e^{\frac{{{\lambda _0} + {\lambda _1}}}{{{\gamma _{{\rm{in}}}}}}}}{{\rm{E}}_{\rm{i}}}\left( { - \frac{{{\lambda _0} + {\lambda _1}}}{{{\gamma _{{\rm{in}}}}}}} \right) - \frac{{\pi {\lambda _1}{e^{\frac{{{\lambda _0}}}{{{\gamma _{{\rm{in}}}}}}}}}}{{8\ln 2}}\zeta \sum\limits_{j = 1}^M {\sqrt {1 - f_j^2} } {e^{ - {\lambda _1}\tan {\theta _j}}}{{\rm{E}}_{\rm{i}}}\left( { - {\lambda _0}\tan {\theta _j} - \frac{{{\lambda _0}}}{{{\gamma _{{\rm{in}}}}}}} \right){\sec ^2}{\theta _j}
\end{align}
\end{small}
\setcounter{equation}{21}
\hrulefill
\vspace*{-4pt}
\end{figure*}
Using the integration by parts with respect to the variable $x$, we can obtain (19). After using the integrations $\int_0^\infty  {{e^{ - \mu x}}\ln \left( {1 + \beta x} \right)dx =  - \frac{1}{\mu }{e^{\frac{\mu }{\beta }}}{{\rm{E}}_{\rm{i}}}\left( { - \frac{\mu }{\beta }} \right)} $ [12, 4.337.2] and $\int_u^\infty  {\frac{{{e^{ - \mu x}}}}{{x + \beta }}dx =  - {e^{\mu \beta }}{{\rm{E}}_{\rm{i}}}\left( { - \mu u - \mu \beta } \right)} $ [12, 3.352.2], (20) can be obtained. First using the change of variables $y=\tan \theta$ and then using the Gaussian-Chebyshev quadrature versus the second term of (21), we can obtain (21). In (21), $\zeta =\frac{\pi}{M}$, ${f_j} = \cos\left( {\frac{{\left( {2j - 1} \right)\pi }}{{2M}}} \right)$, ${\theta_j} = \frac{\pi}{4}({f_j} + 1)$, and  $M$ is the parameter determining the tradeoff between the calculating complexity and the accuracy.

Substituting $\gamma_{\rm{op}}^2$ into $C_{2}$, $C_{2}$ can be derived as (22)$-$(25), which are detailed at the top of next page. Step $(b_{1})$ utilizes the Gaussian-Chebyshev quadrature with respect to the  variable $x$, and step $(b_{2})$ and $(b_{3})$ use the Gaussian-Chebyshev quadratures with respect to the variables $y$ and $z$, after changing the variables of $y=\tan\theta$ and $z=\tan\theta$, respectively. In (23)$-$(25), ${\omega _1}=\frac{\pi}{N_{1}}$, ${\omega _2}=\frac{\pi}{N_{2}}$, ${\omega _1}=\frac{\pi}{N_{3}}$, $f_{i_{1}}=\cos(\frac{(2i_{1}-1)\pi}{2N_{1}})$, ${c_{{i_1}}} = {f_{{i_1}}} + 1$, $f_{i_{2}}=\cos(\frac{(2i_{2}-1)\pi}{2N_{2}})$, ${\theta _{{i_2}}} = \frac{\pi}{4}(f_{i_2} + 1)$,
$f_{i_{3}}=\cos(\frac{(2i_{3}-1)\pi}{2N_{3}})$, ${\theta _{{i_3}}} = \frac{\pi}{4}(f_{i_3} + 1)$, $\Phi  = \frac{{{\lambda _0}{\lambda _1}\pi {\omega _1}}}{{16}}$ and ${\Phi _{{i_1}{i_2}{i_3}}} = \frac{{{\lambda _0}{\lambda _1}{\lambda _2}{\pi ^2}}}{{64}}{\omega _1}{\omega _2}{\omega _3}\sum\limits_{{i_1} = 1}^{{N_1}} {\sqrt {1 - f_{{i_1}}^2} } \sum\limits_{{i_2} = 1}^{{N_2}} {\sqrt {1 - f_{{i_2}}^2} } \sum\limits_{{i_3} = 1}^{{N_3}} {\sqrt {1 - f_{{i_3}}^2} } $. $N_1$, $N_2$ and $N_3$ are the parameters determining the tradeoff between the calculating complexity and the accuracy.
\newcounter{mytempeqncnt1}
\begin{figure*}[t]
\normalsize
\setcounter{mytempeqncnt}{\value{equation}}
\setcounter{equation}{21}
\begin{small}
\begin{align}
%\begin{array}{l}
%\[\begin{array}{l}
%\[\begin{array}{l}
{C_2} &= \frac{1}{2}\int_0^\infty  {\int_0^\infty  {\int_0^y {{\lambda _0}{e^{ - {\lambda _0}x}}{\lambda _1}{e^{ - {\lambda _1}y}}{\lambda _2}{e^{ - {\lambda _2}z}}{{\log }_2}\left( {1 + {\gamma _{{\rm{in}}}}\frac{{x + \eta yz}}{{1 + \eta z}}} \right)dxdydz} } } \\
&\mathop {{\rm{  }} \approx }\limits^{({b_1})} \frac{1}{2}\int_0^\infty  {{\lambda _2}{e^{ - {\lambda _2}z}}\int_0^\infty  {{\lambda _1}{e^{ - {\lambda _1}y}}} } \left( {\sum\limits_{{i_1} = 1}^{{N_1}} {\frac{y}{2}{\omega _1}\sqrt {1 - f_{{i_1}}^2} } {\lambda _0}{e^{ - {\lambda _0}\left( {\frac{y}{2}{c_{{i_1}}}} \right)}}{{\log }_2}\left( {1 + {\gamma _{{\rm{in}}}}\frac{{\frac{y}{2}{c_{{i_1}}} + \eta yz}}{{1 + \eta z}}} \right)} \right)dydz\\
&\mathop  \approx \limits^{({b_2})} \Phi \sum\limits_{{i_1} = 1}^{{N_1}} {\sqrt {1 - f_{{i_1}}^2} } \left( {\int_0^\infty  {{\lambda _2}{e^{ - {\lambda _2}z}}\tan {\theta _{{i_2}}}{\omega _2}\sum\limits_{{i_2} = 1}^{{N_2}} {\sqrt {1 - f_{{i_2}}^2} } } {e^{ - \left( {\frac{{{\lambda _0}{c_{{i_1}}}}}{2} + {\lambda _1}} \right)\tan {\theta _{{i_2}}}}}{{\sec }^2}{\theta _{{i_2}}}{{\log }_2}\left( {1 + {\gamma _{{\rm{in}}}}\frac{{\frac{{{c_{{i_1}}}}}{2}\tan {\theta _{{i_2}}} + \eta z\tan {\theta _{{i_2}}}}}{{1 + \eta z}}} \right)dz} \right)\\
&\mathop  \approx \limits^{(b3)} {\Phi _{{i_1}{i_2}{i_3}}}\left( {{e^{ - \left( {\frac{{{\lambda _0}{c_{{i_1}}}}}{2} + {\lambda _1}} \right)\tan {\theta _{{i_2}}}}}{e^{ - {\lambda _2}\tan \left( {{\theta _{{i_3}}}} \right)}}\tan {\theta _{{i_2}}}{{\sec }^2}{\theta _{{i_2}}}{{\sec }^2}{\theta _{{i_3}}}{{\log }_2}\left( {1 + {\gamma _{{\rm{in}}}}\frac{{\frac{{{c_{{i_1}}}}}{2}\tan {\theta _{{i_2}}} + \eta \tan {\theta _{{i_2}}}\tan {\theta _{{i_3}}}}}{{1 + \eta \tan {\theta _{{i_3}}}}}} \right)} \right)
%\end{array}\]
%\end{array}\]
%\end{array}
\end{align}
\end{small}
\setcounter{equation}{25}
\hrulefill
\vspace*{-4pt}
\end{figure*}
%\vspace{-15pt}
\vspace{-5pt}
\section{Numerical Results and Discussion}
This section presents the numerical results to validate above analysis. The parameters are set to as follows in the simulations [9]: $\eta=0.5$, $\frac{1}{\lambda_{0}}=1$, $\frac{1}{\lambda_{1}}=5$ and  $\frac{1}{\lambda_{2}}=5$.

\begin{figure}[htbp]
\centering
\includegraphics[width=2.8 in]{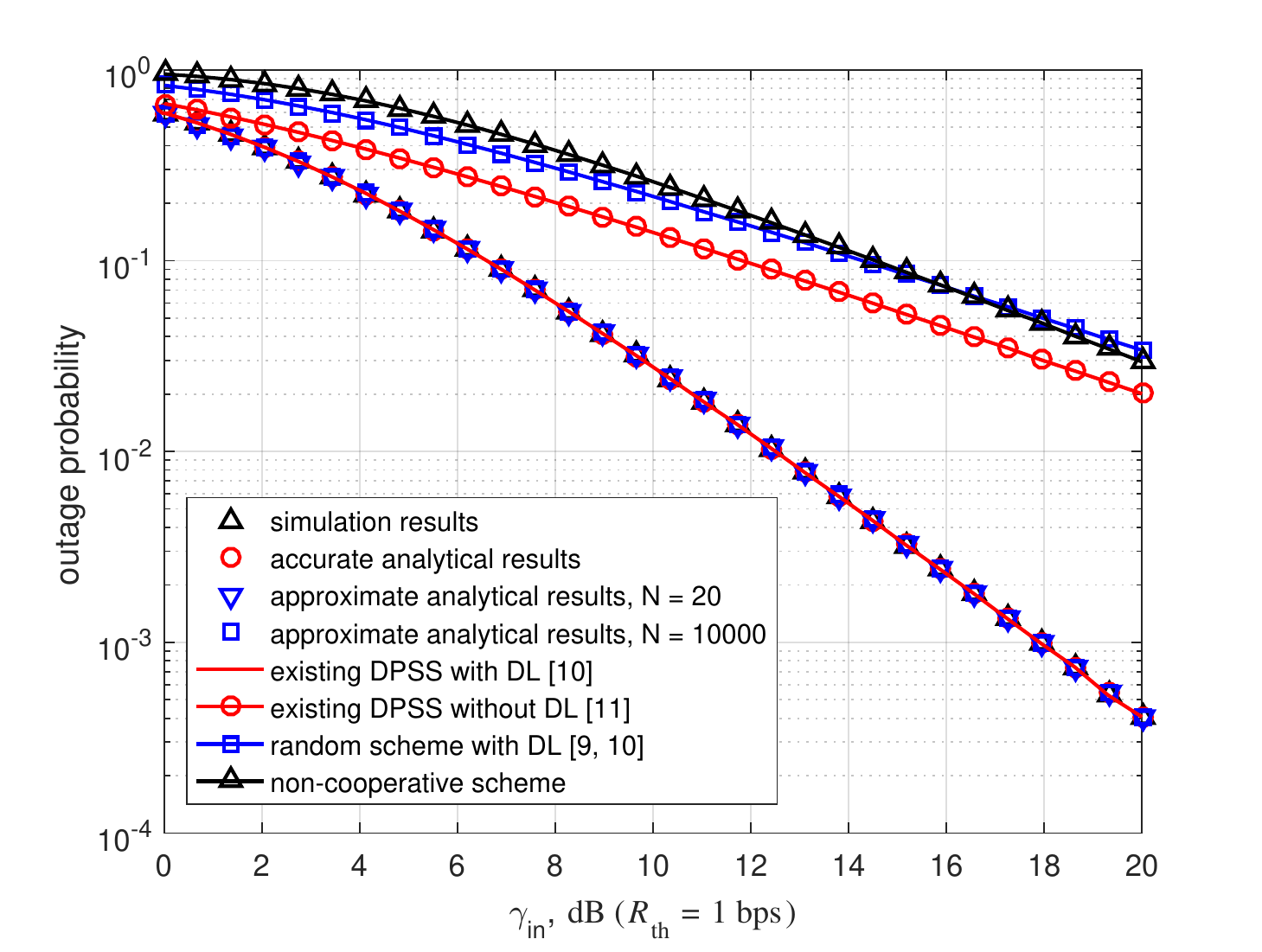}
\caption{The comparisons of the proposed DPSS and the existing schemes on the outage probability.}
\label{Fig.4}
\end{figure}
As shown in Fig.1, both the approximate and accurate results both match well with the simulation results over the entire SNR regime for the proposed scheme. The outage probability of the proposed DPSS is smaller than that of the existing DPSS without DL and the random scheme with DL, and compared to the non-cooperative scheme, the DL indeed improve the performance in the proposed DF based cooperative SWIPT network.

In the existing DPSS in [6, 10], the dynamic PS factor is selected based on the target data rate of the system and the instantaneous CSI and can be written as ${\rho ^*}{\rm{ = }}\max \left( {1 - \frac{{{\gamma _{{\rm{th}}}}}}{{{\gamma _{{\rm{in}}}}{{\left| {{h_1}} \right|}^2}}},0} \right)$ [10, eq.16]. If the selected dynamic PS factor in [6, 10] cannot support the target rate, no other dynamic PS factor can accomplish the transmission at the target rate in the proposed DPSS. So, two scheme have the same performance in terms of outage probability which is verified in Fig.1.
\begin{figure}[htbp]
\centering
\includegraphics[width=2.6 in]{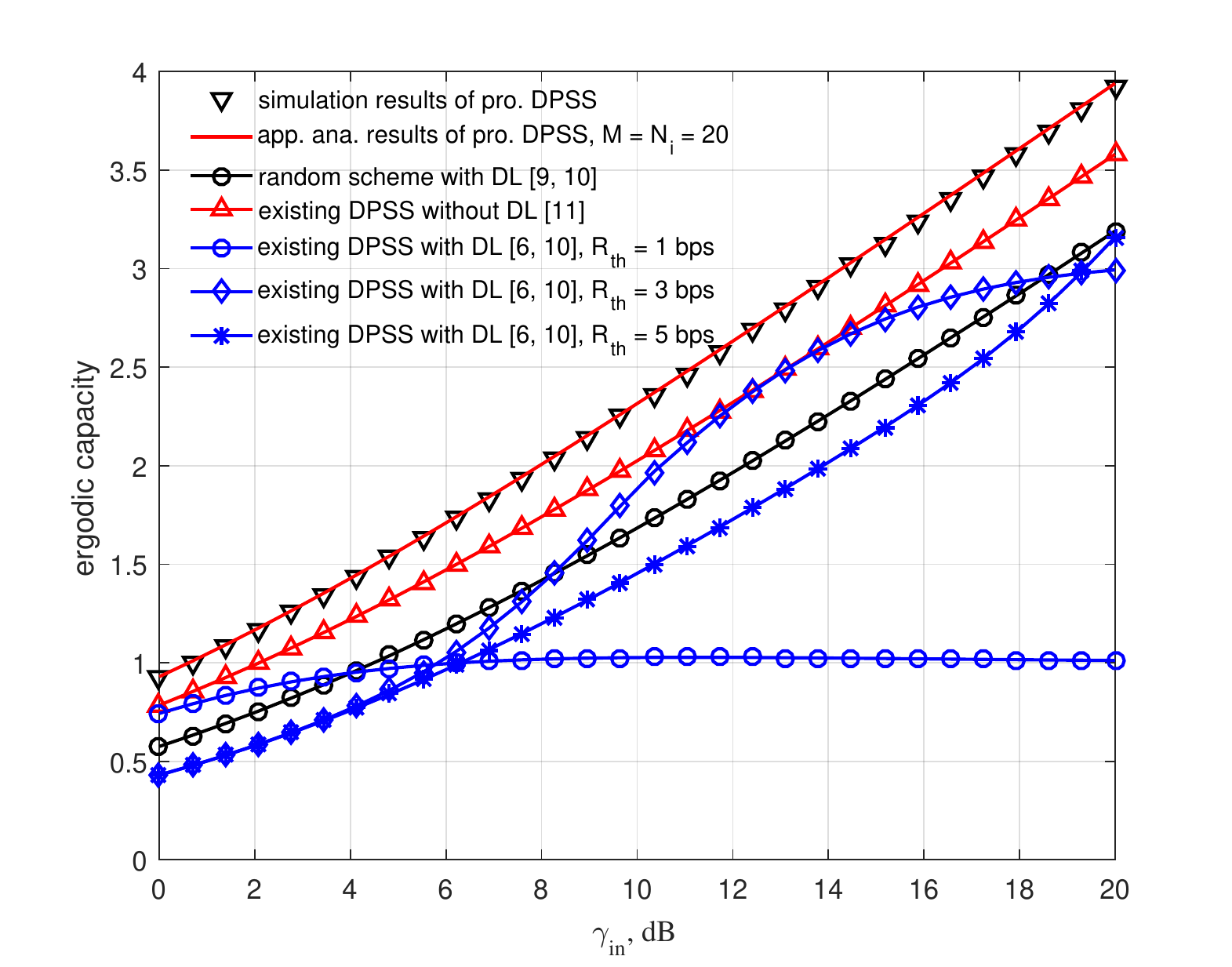}
\caption{The comparisons of the proposed DPSS and the existing schemes on the ergodic capacity. ($M=N_{1}=N_{2}=N_{3}=20$).}
\label{Fig.5}
\end{figure}

As shown in Fig.2, the proposed DPSS can maximize the received SNR at the destination and utilize the DL, so the proposed scheme has the highest ergodic capacity compared to other existing schemes over the entire SNR regime.
%\vspace{-15pt}
\section{Conclusions}
In this letter, a DPSS for the DF based cooperative SWIPT
network with direct link has been proposed. The proposed DPSS can minimize the outage probability and maximize the ergodic capacity simultaneously. Numerical results showed that the proposed scheme has the best performance compared to other existing schemes.
%\vspace{-15pt}
\appendices


\begin{thebibliography}{1}

%\bibitem{Yi}
%Z. Yi, M. Ju and I. Kim, ``Outage Probability and Optimum Combining for Time Division Broadcast Protocol,'' \emph{IEEE Transactions on Wireless Communications}, vol. 10, no. 5, pp. 1362-1367, May 2011.

\bibitem{Perera}
T. D. P. Perera, D. N. K. Jayakody, S. K. Sharma, S. Chatzinotas, and
J. Li, ``Simultaneous wireless information and power transfer (SWIPT):
Recent advances and future challenges,'' \emph{IEEE Commun. Surveys Tuts.},
vol. 20, no. 1, pp. 264-302, 1st Quart., 2018.

\bibitem{Nasir}
A. A. Nasir, X. Zhou, S. Durrani, and R. A. Kennedy, ``Relaying protocols
for wireless energy harvesting and information processing,'' \emph{IEEE Trans.
Wireless Commun.}, vol. 12, no. 7, pp. 3622-3636, Jul. 2013.

\bibitem{Nasir}
A. A. Nasir, X. Zhou, S. Durrani, and R. A. Kennedy, ``Throughput
and ergodic capacity of wireless energy harvesting based DF relaying
network,'' in  \emph {Proc. IEEE ICC}, Jun. 2014,
pp. 4066-4071.

\bibitem{Ye}
Y. Ye, L. Shi, X. Chu, H. Zhang, and G. Lu, ``On the Outage Performance of SWIPT Based Three-step Two-way DF Relay Networks,'' \emph{ IEEE Trans. on Veh. Technol.}.
doi: 10.1109/TVT.2019.2893346

\bibitem{Do}
T. P. Do, I. Song, and Y. H. Kim, ``Simultaneous wireless transfer
of power and information in a decode-and-forward two-way
relaying network,'' \emph{IEEE Trans. Wireless Commun.}, vol. 16, no. 3,
pp. 1579-1592, Mar. 2017.

\bibitem{Ding}
Z. Ding, S. M. Perlaza, I. Esnaola and H. V. Poor, ``Power Allocation Strategies in Energy Harvesting Wireless Cooperative Networks,'' \emph{IEEE Trans.
Wireless Commun.}, vol. 13, no. 2, pp. 846-860, February 2014.

\bibitem{Laneman}
J. N. Laneman, D. N. C. Tse, and G. W. Wornell, ``Cooperative
diversity in wireless networks: Efficient protocols and outage behavior,''
\emph{IEEE Trans. Inf. Theory}, vol. 50, no. 12, pp. 3062-3080, Dec. 2004.

\bibitem{Kumar}
P. Kumar and K. Dhaka, ``Performance Analysis of Wireless Powered DF Relay System Under Nakagami-$m$ Fading,'' \emph{IEEE Trans. Veh. Technol.}, vol. 67, no. 8, pp. 7073-7085, Aug. 2018.

\bibitem{Lee}
H. Lee, C. Song, S. H. Choi, and I. Lee, ``Outage probability analysis and
power splitter designs for SWIPT relaying systems with direct link,'' \emph{IEEE
Commun. Lett.}, vol. 21, no. 3, pp. 648-651, Mar. 2017.

\bibitem{Ye}
Y. Ye, Y. Li, F. Zhou, N. Al-Dhahir, and H. Zhang, ``Power Splitting-Based SWIPT With Dual-Hop DF
Relaying in the Presence of a Direct Link,''  \emph{IEEE Syst. J.}, doi: 10.1109/JSYST.2018.2850944.

\bibitem{Ashraf}
M. Ashraf, et al., ``Capacity Maximizing Adaptive Power Splitting Protocol for Cooperative Energy Harvesting Communication Systems,'' \emph{IEEE
Commun. Lett.}, vol. 22, no. 5, pp. 902-905, May 2018.


\bibitem{Gradshteyn}
I. S. Gradshteyn and I. M. Ryzhik, Table of Integrals, Series, and Products,
7th ed. London, U.K.: Academic, 2007.

\end{thebibliography}
\end{document}